\documentclass{llncs}

\usepackage[english]{babel}
\usepackage{euscript}
\usepackage{amssymb}
\usepackage{amsmath}
\usepackage{makeidx}
\usepackage{mathabx}
\usepackage{pslatex}
\usepackage[T1]{fontenc}
\usepackage{proof}
\usepackage{stmaryrd}
\usepackage{mathrsfs}
\usepackage{subfigure}

\newcommand{\act}{\vartriangleright}
\newcommand{\mess}{\vartriangleleft}
\newcommand{\FN}{\mathcal{FN}}

\newcommand{\FV}{\mathcal{FV}}

\newcommand{\ie}{\textit{i.e.}~}
\newcommand{\Ie}{\textit{I.e.}~}
\newcommand{\cf}{\textit{cf.}~}
\newcommand{\etal}{~\textit{et al.}~}

\newtheorem{thm}{Theorem}

\makeatletter
\def\acksec{%
  \@startsection
    {section}{1}{\z@}{18\p@ \@plus 2\p@ \@minus 2\p@}%
    {6\p@}{\normalfont\large\bf}}

\makeatother

\begin{document}

\title{Static Safety for an Actor Dedicated Process Calculus by Abstract Interpretation}
\titlerunning{Static Safety for Actors by Abstract Interpretation}
\author{Pierre-Loïc Garoche \and Marc Pantel \and Xavier Thirioux}
\authorrunning{Garoche \etal}

\institute{IRIT, Toulouse\\
\email{$\{$garoche,pantel,thirioux$\}$@enseeiht.fr} %,\\ 
%\texttt{http://www.irit.fr/~Pierre-Loic.Garoche}
}

\maketitle
\begin{abstract}
The actor model eases the definition of concurrent programs with non uniform behaviors. Static analysis of such a model was previously
done in a data-flow oriented way, with type systems. This approach was based on constraint set resolution and was not able to deal 
with precise properties for communications of behaviors. 
We present here a new approach, control-flow oriented, based on the abstract interpretation framework, able to
deal with communication of behaviors. 
Within our new analyses, we are able to verify most of the previous properties we observed as well as new ones, principally based on 
occurrence counting.
\end{abstract}

\section{Introduction}
\subsection{Context -- Motivation}
 The development of the telecommunication industry and the generalization of
network use bring concurrent and distributed programming in the limelight. In
that context, programming is a hard task and, generally, the resulting
applications contain much more \emph{bugs} than usual centralized software. As
sequential object oriented programming is commonly accepted as a \emph{good} way
to build software, concurrent object oriented programming seems to be
well-suited for programming distributed systems.
%%%%
Since non-determinism resulting from network communications makes it
difficult to validate any distributed functionality using informal approaches, our work
is focused on applying formal methods to improve concurrent
object oriented programming.

To obtain widely usable tools, we have chosen to use the actor model proposed by
\textsc{Hewitt}~\cite{HBS73a} and developed by \textsc{Agha}~\cite{A86a}. This model is
based on a network of autonomous and cooperative agents (called actors), which
encapsulate data and programs, communicating using an asynchronous point to
point protocol.  An actor stores each received message in a queue and when
idle, processes the first message it can handle in this queue.
%%%%
Besides those conventions (which are also true for concurrent objects), an actor
can dynamically change its interface. This property 
allows to increase or decrease the set of messages an actor may handle, yielding
a more accurate programming model. This model, also known as concurrent
objects with non uniform behavior (or interface), has been adopted by the
telecommunication industry for the development of distributed and concurrent
applications for the Open Distributed Computing framework (ITU X901-X904) and
the Object Description Language (TINA-C extension of OMG IDL with multiple
interfaces).
%%%%
Until now, we have been designing several analyses for an actor model, all of which 
based on typing systems. Our main objective was, and still is, to detect in a most accurate way 
typical flaws of distributed applications, like for instance communication deadlock or non linearity 
(i.e. the fact that several distributed actors have the same address).
%%%%
Due to limitations of our previous attempts, which we could somehow overcome 
but at the price of a much greater complexity unmatched with only a small gain in precision, 
we decided to move to the framework of abstract interpretation, whose tools and ideas have now significantly
 grown in maturity and are being 
widely used in industrial contexts, or are on the verge of being so. We now investigate 
these techniques in order to capture our long standing properties of interest
(detection of orphan messages, that is messages sent to an actor which will not handled them)
 as well as new ones, especially dedicated to control of resources' usage.

In a first section, we define our actor calculus. Then in the second part, we introduce our non standard semantics upon which 
we define, in the third part, an abstraction. Finally, in the last part, we explain how to use the abstraction to observe properties
about an analyzed term.

\subsection{Related works}
Concerning concurrent objects and actors with uniform or non-uniform
behaviors, and more generally process calculi, typing systems (usually related to data-flow like analysis) have been the subject of active research. 
Two opposite approaches have been followed: type declaration and type inference. In the first case, most proposals make use of types as processes of a simple algebra,
for instance CCS (Calculus of Communicating Systems) processes. This allows a form of subtyping through simulation relations or language containment. 
The works of  \textsc{Kobayashi}\etal\cite{kobayashi04,kobayashi02}, \textsc{Ravara}\etal\cite{ravara.vasconcelos:typing-non-uniform-objects}, \textsc{Najm}\etal\cite{carrez03behavioural,NNS99a}, \textsc{Puntigam}~\cite{puntigam96types}, and \textsc{Hennessy}\etal\cite{HRY04-slcmc}
follow this line of thought, to which we can add the works of \textsc{Rajamani}\etal\cite{chaki02types,rajamani01behavioral}, bringing model-checking issues for those processes-as-types
in the scope. The second case is again twofold: on one side we have unification based typing algorithms focusing on resources' usage control witnessed by the works of \textsc{Fournet}\etal\cite{FLMR97a} and 
\textsc{Boudol}\etal\cite{boudol:typing-use}, whereas on the other side we have flow based algorithms, related to behavior and communication patterns reconstruction, advocated by the works of 
\textsc{Nielson}\etal\cite{amtoft97type} and \textsc{Pantel}\etal\cite{CoPaDaSa1999.1,CoPaSa2000.1,CoThPa2003.1}. 
Explicit typing may provide more precise information but are sometimes very hard to write
for the programmer (they might be much more complex than the program itself).
Implicit typing requires less user supplied information but lead to less precise results. 

One drawback of type-based analyses is that they are mainly concerned with data-flow analyses (as types basically represent
sets of possible values for variables). In this context, control flow analyses can be mimicked with sophisticated encodings~\cite{palsberg95} but abstract interpretation
seems to be more adequate in this respect. It has been recently applied with success to concurrent and distributed programming 
by the work of \textsc{Venet}~\cite{venet:phd} and later \textsc{Feret}~\cite{feret:jlap'05,feret_eng:phd}.

\section{CAP: a primitive actor calculus}
\label{cap}

In order  to ease the definition of static analysis for actor based programming, we proposed, in 96, 
the CAP primitive actor calculus~\cite{CoPaSa1996}, 
which merge asynchronous $\pi$-calculus and \textsc{Cardelli}'s Primitive Object Calculus.
The following example illustrates both replication and behavior passing mechanisms of CAP. 
The $\nu$ operator defines two addresses, $a$ and $b$, then two actors denoted by program points
$1$ and $7$ are defined on those addresses with the behavior set respectively denoted by $2$ and $4$ for $a$ and $8$ for $b$.

At this point the actor $1$ can handle messages called $m$ or $send$ when $b$ can only handle $beh$ messages.
\begin{center}
\(
\begin{array}{r l}
\nu a^\alpha, b^\beta, & a \act^1 [m^2() = \zeta(e,s)(a \act^3 s), \\
& ~~~~~~~send^4(x) = \zeta(e,s)(x \mess^5 beh(s))] \\
~||~ & a \mess^6 send(b) \\
~||~ & b \act^7 [beh^8(x) = \zeta(e,s)(e \act^9 x)] \\
~||~ & b \mess^{10} m()
\end{array}
\)
\end{center}
There are also two messages in the initial configuration. One is labeled $send$ and is sent to $a$, the other one is labeled $m$
and is sent to $b$. In the initial configuration, there is only one possible interaction, in which the actor 
$a$ handles the message
$send$. The message $m$ is an orphan one: it is in the configuration but cannot be handled for the moment.
After one interaction between $a$ and the message $send$, the message $beh$ which argument is the behavior's set of $a$ is sent
to $b$. Thus $b$ can handle that message. In its continuation, the actor $b$ assumes the behavior's set
of $a$. Thus $b$ can now handle the message $m$. This example shows how to send a behavior to another actor. Such a mechanism 
increases the difficulty of statically inferring properties. 
Stuck-freeness, \ie the detection of the set of permanent orphans messages, or
linearity, \ie verifying that at most one actor is associated to a particular address at the same time, are harder to statically
infer when we allow behavior passing. This point was one of the constraints which led us to switch from type based analysis to
abstract interpretation.

\subsection{Syntax and semantics}
Let $\mathscr{N}$ be an infinite set of actor names, $\mathscr{V}$ be an infinite set of variables.
Let $\mathscr{L}_m$ be a set of message labels, $\mathscr{L}_p$ be the set of
program point labels and $\mathscr{L}_n$ be the set of name labels. 
In the following, we denote $\mathscr{L}_p \cup \mathscr{L}_n$ by $\mathscr{L}$.
The syntax of configurations is described as follows:
\begin{center}
\(
\begin{array}{c c l}
C &::=& 0 ~~|~~ \nu a^{\alpha} ~ C ~~|~~ C ~||~ C ~~|~~ a \act^{l} P ~~|~~ a \mess^l m(\widetilde{P}) \\
%&&\\
P &::=& x \qquad | \qquad [{m_i^{l_i}(\widetilde{Var}) = \zeta(e,s) C_i}^{i=1\ldots n} ]\\
\end{array}
\)
\end{center}
Configurations can be an empty process, a creation of actor's address, parallel execution, an actor on address $a$ with 
behavior defined by $P$ and, finally, a message sent to an address $a$ with arguments $\widetilde{P}$.
Program points define messages, behaviors' installation or external choices between some actors' behaviors. They will be
used to build traces of the execution control flow.
Name restriction, in the configuration $(\nu a^\alpha)C$, acts as a name binder, 
so does the $\zeta$ operator and the message label
for variables in the behavior description of an actor, \ie 
in the behavior $[{m_i^{l_i}(\widetilde{x_i}) = \zeta (e_i,s_i) C_i}^{i=1\ldots n}]$ , therefore
the occurrences of $a$ in $C$, $\widetilde{x_i}$ in $\zeta(e_i,s_i) C_i$ and $e_i$ and $s_i$ in $C_i$ are bound.
The $\zeta$ operator is our reflexivity operator, it catches both address and behavior of its actor and allows to re-use them in
the behavior.
We denote by $\FN(C)$ the set of free names in $C$ and by $\FV(C)$ the set of free variables.
The standard semantics of CAP was defined, à la Milner, by both the usual transition rule (\cf Fig.~\ref{trans_rule}) 
and the congruence relation (\cf Fig.~\ref{cong_rel}).

\begin{figure}
\[
\infer[~]{
  a \act T  ~||~ a \mess^l m(\widetilde{T_l})
  \xrightarrow[ ]{\textrm{comm}(l,l_k)}
  C_k[e_k \leftarrow a, s_k \leftarrow T, \widetilde{x_k} \leftarrow \widetilde{T_l}]}
  {
    T = [{m_i^{l_i}(\widetilde{x_i}) = \zeta (e_i,s_i) C_i}^{i=1,\ldots, n} ]
    &
    \left\{ \begin{array}{l c l} m = m_k,\\ length(\widetilde{T_l}) = length(\widetilde{x_k}),\\ k \in [1, \ldots,n] \end{array} \right.}
\]
In order to distinguish transitions, we label the interacting parts of terms. Here the message has label $l$ and the matching behavior
label $l_k$.
\caption{Transition rule of CAP standard semantics}
\label{trans_rule}
\end{figure}

\begin{figure}[t]
\[\begin{array}{c}
\begin{array}{c c c r l}
C &\equiv& D & \textrm{C }  \alpha \textrm{-convertible} \textrm{ to D} & (\alpha -conversion)\\
C || 0 &\equiv& C && (inaction) \\
C || D &\equiv& D || C && (commutativity) \\
(C || D) || E &\equiv& C || (D || E) && (associativity) \\
(\nu a) \varnothing & \equiv& \varnothing & (garbage~collecting)\\
T \act T_1 &\equiv& T \act T_2 & \textrm{if } T_1 \equiv T_2 & (behavior~equivalence)\\
(\nu a)(\nu b)C &\equiv& (\nu b)(\nu a)C & \textrm{if }a \neq b & (swapping) \\
(\nu a)C || D &\equiv& (\nu a)(C || D) & \textrm{if } a \notin \FN(D) & (extrusion)
\end{array}
\end{array}
\]
\caption{Congruence relation of CAP standard semantics}
\label{cong_rel}
\end{figure}

\section{Non standard semantics}

In order to ease the definition of abstract interpretations, we need to define 
define, in this section, another semantics for CAP and prove it 
bisimilar to standard CAP semantics. The non standard semantics allows
us to label each process with the history of transitions which led to both its creation and the creation of its values. 
Our work is based on a generic non standard semantics which has been defined by \textsc{Feret}~\cite{feret:jlap'05,feret_eng:phd}
to model first order process calculi as $\pi$-calculus,
%~\cite{milner93polyadic}
spi-calculus,
%~\cite{abadi97calculus}
Ambients,
%~\cite{cardelli98mobile}
Bio-ambients
%~\cite{BioAmbientsTCS} and Join 
calculus.
%~\cite{fournet96reflexive}
We also describe in this section how we adapt this general framework to
express the CAP language which has a notion of higher order due to its behavior passing and reflexivity mechanism ($\zeta$ operator).
We then briefly 
describe the operational semantics of the generic non standard semantics.

A configuration of a system, in this semantics, is a set of threads. Each thread $t$ is a triple defined as
\(t = (p,id,E) \in \mathscr{L}_p \times \mathscr{M} \times (\mathscr{V} \mapsto (\mathscr{L} \times \mathscr{M})) \) where 
$p$ is the program point
representing the thread in the CAP term, 
$id$ is the history marker, also called its identity, and $E$ its environment.
This environment is a partial map from a variable to a pair $(value, marker)$. Each marker is a word on program points
representing the history of 
transitions which led to the creation of values or threads. 
It is required in order to differentiate recursive instances of a value or thread. All threads with
the same program point have an environment defined on the same domain, called the program point interface.

We will describe some primitives that allow us to define the non standard semantics, then, briefly, we show how to compute 
transitions in this semantics.

\subsection{Partial interactions}
We associate to each program point a partial interaction which defines how threads related to this program point can interact with
others. We also define the set of variables associated to each thread, constituting its environment, according to its program point.
Here, in CAP, partial interactions can represent a syntactically defined actor, a dynamic one (an actor whose behavior is 
defined by a variable) and a particular behavior of an actor or a sent message.

We thus define the set of partial interactions names 
$\mathscr{A} = \{static\_actor_{n} , behavior_{n},$ $message_{n} ~|~ n \in \mathbb{N} \} \cup \{ dynamic\_actor \} $
and their arities as follows:
\begin{center}
\(
\begin{array}{l c l}
Ari &=& \left\{
static\_actor_n  \mapsto (2,n), 
dynamic\_actor  \mapsto (2,0), 
behavior_n \mapsto (1,n+2), \right.\\
& & \left. message_n \mapsto (n+2,0)
\right\}
\end{array}
\)
\end{center}
Partial interaction arities define the number of parameters and the number of bound variables.

The partial interaction \textit{$dynamic\_actor$} denotes a thread representing an actor. It is consumed when interacting.
It has only two parameters: its name and set of behaviors. It binds no variables.

Both partial interaction \textit{$static\_actor_n$} and \textit{$behavior_n$} denote a particular behavior of an actor. 
The first one is associated to an address when the second one is alone and can be used with a dynamic actor.
The second one acts as a definition and stays in the configuration when used, whereas the first one is deleted.
They are parametrized by their message labels and binds $n+2$ variables, the variables under the $\zeta$
operator expressing reflexivity as well as the parameters of the message it can handle.
The first one is also parametrized by its actor's name.

Finally the partial interaction \textit{$message_n$} represents the message that is sent to a particular address (actor).
So it has $n+2$ parameters: one for the address, one for the message name and $n$ for the variables of this message. 
It is consumed when interacting.

We associate to each partial interaction a type denoting whether such a partial interaction is consumed or not when interacting.

\subsection{Abstract syntax extraction}
\label{ase}
We now define the syntax extraction function that takes a CAP term describing the initial state of an agents'
system in the standard syntax and extracts its abstract syntax.

We map each program point labeled $l \in \mathscr{L}_p$ to a set of partial interaction and to an interface.

A partial interaction $pi$ is given by a tuple $(s, (parameter_i), (bound_i), cons\-traints, con\-tinuation)$
where $s \in \mathscr{A}$ is a partial interaction name, $(m,n) = Ari(s)$ its arity,
$(parameter_i) $ $\in \mathscr{V}^m$ its finite sequence of variables ($X_i$), $(bound_i) \in \mathscr{V}^n$ its finite sequence of
distinct variables ($Y_i$), \(constraints \subseteq \{ v \diamond v' ~|~ (v,v') \in \mathscr{V}^2, \diamond \in \{ =, \neq \} \}\)
its synchronization constraints and finally $continuation \in \wp(\mathscr{L}_p \times\-(\mathscr{V} \rightarrow \mathscr{L}))$
its syntactic continuation. We will check constraints defined in the set $constrains$ about thread environment with 
the use of the sequence $(parameter_i)$, then we will 
use both sequences $(parameter_i)$ and $(bound_i)$ to compute value passing, finally we will deal with the set $continuation$ to 
determine which threads have to be inserted in the system.

\begin{itemize}

\item the label of a program point $a \act^l [{m_i^{l_i}(\widetilde{x_i}) = \zeta (e_i,s_i) C_i}^{1 \leq i \leq m}]$ 
is associated to the interface $\{ a \}$ and to the following
set of partial interactions:
\[ \left\{ \begin{array}{l}
\Big\{ ( static\_actor_n , [a,m_1], [e_1,s_1,\widetilde{x_1}], \beta(C_1,\emptyset) ) \Big\} \\
\Big\{ ( static\_actor_n , [a,m_2], [e_2,s_2,\widetilde{x_2}], \beta(C_2,\emptyset) ) \Big\} \\
\ldots \\
\Big\{ ( static\_actor_n , [a,m_m], [e_m,s_m,\widetilde{x_m}], \beta(C_m,\emptyset) ) \Big\} \\
\end{array} \right\} \]

\item the label of a program point $a \act^l x $ is associated to the interface $\{ a,x \}$ and to the following
set of partial interactions:
\(\Big\{ ( dynamic\_actor , [a,x],  \emptyset , \emptyset ) \Big\}\)

\item the label of a program point $a \mess^l m(\widetilde{P})$ is associated to the interface
$\{ a\} \cup \FV(\widetilde{P})$ and to the following set of
partial interactions:
\(\Big\{ ( message_n , [a;m; \widetilde{P} ] ,\) \( \emptyset , \emptyset ) \Big\}\)

\item the label of a program point $l_i$ corresponding to a particular behavior of an actor \ie
  $m_i^{l_i}(\widetilde{x}) = \zeta (e_i,s_i) C_i$ is associated to
  the interface $\FV(C_i) \setminus \{ e_i,s_i\} $ and
  to the following set of partial interactions:
  \(\Big\{ ( behavior_n , [m_i], [e_i,s_i,\widetilde{x}], \)\ \( \beta(C_i,\emptyset) ) \Big\}\)

\end{itemize}
Finally, the syntax extraction function $\beta$ is defined inductively over the standard syntax of the syntactic 
continuation, as follows:
\begin{center}
\(
\begin{array}{r c l}
\beta ((\nu a^\alpha)C,E_s) &=& \beta (C,E_s[a \mapsto \alpha])\\
%%
%\\
\beta (\emptyset, E_s) &=& \{ \emptyset \}\\
%%
%\\
\beta (C_1 || C_2, E_s) &=& \beta(C_1,E_s) \cup \beta(C_2,E_s) \\
%%
%\\
%\begin{array}{r}
\beta (a \act^l [m_i^{l_i}(\widetilde{x_i}) =
%\\
{\zeta(e_i,s_i) C_i}^{i=1,\ldots, n}], E_s)
%\end{array}
&=&
\left\{ (l,E_s) \right\} 
\cup  \bigcup_{i=1,\ldots,n} \left\{
(l_i,E_s)
\right\}
\\
%%
%\\
\beta (a \act^l B, E_s) &=& \{ \{ (l,E_s) \} \}\\
%\\
\beta (a \mess^l m(\widetilde{P}), E_s) &=& \{ \{ (l,E_s) \} \}
\end{array}
\)
\end{center}

The initial state for a term $\mathscr{S}$ is described by $init_s$, a set of potential continuations in 
$\wp(\wp(\mathscr{L}_p \times (\mathscr{V} \rightarrow \mathscr{L})))$ defined as $\beta(\mathscr{S},\emptyset)$.

\subsection{Formal Rules}
\label{fr}

We now define the formal rules that drive the interaction between threads.
In the case of CAP, we have two rules that describe an actor handling a message, depending on the kind of actor we have, a static or a
dynamic one.

In the following, the $i$-th parameter, the $j$-th bounded variable, and the identity of the $k$-th partial interaction are
respectively denoted by $X_i^k$, $Y_j^k$ and $I^k$. 
\label{behavior_set_function}
We define the endomorphism $behavior\_set$ on the set 
$\mathscr{L}_p \times \mathscr{M}$ as follows: $(p,m) \mapsto (p',m)$ where $p$ is a behavior program point and
$p'$ is the program point where $p$ has been syntactically defined.
As an example, in the term $\nu^\alpha a, a \act^1 [m^2() = \zeta (e,s) C]$, we have $behavior\_set(2,m) = (1,m)$.

\paragraph{Communication with a syntactic defined actor.}
The first rule needs two threads, the first one must denote a partial interaction $static\_actor$ when the second one must
denote a partial interaction $message_n$. We both check that the actor's address ($X_1^1$) is equal to the message's 
receiver ($X_1^2$) and that the actor behavior label ($X_2^1$) is equal to the message label ($X_2^2$).

We then define $v\_passing$ that describe the value passing due to both the $\zeta$ operator and message handling. 
%\begin{figure}[h!]
\begin{center}
\(
static\_trans_n = (2, components , compatibility , v\_passing )
\)
\end{center}
where\\
\begin{tabular}{l l}
%\begin{enumerate}
%\item 
1. $components  = \left\{ \begin{array}{l c l} 
    1 &\mapsto& static\_actor_n,\\ 
    2 & \mapsto& message_n
  \end{array}
  \right.$
&
%\item 
2. $compatibility = \left\{ \begin{array}{l}
  X_1^1 = X_1^2;\\
  X_2^1 = X_2^2 ; \\
\end{array} \right. $
\\ \\
%\item 
3. $v\_passing = \left\{ \begin{array}{l}
Y_1^1 \leftarrow X_1^1;\\
Y_2^1 \leftarrow I^1;\\
Y_{i+2}^1 \leftarrow X_{i+2}^2, \forall i \in \llbracket 1;n \rrbracket;
\end{array}
\right.$
%\end{enumerate}
\end{tabular}
%\caption{Communication with a syntactic defined actor}
%\label{rule_1}
%\end{figure}

\paragraph{Communication with a dynamic actor.}
The second rule needs three threads: the first one must denote a partial interaction $behavior_n$, the second one a partial interaction
$dynamic\_actor$ and the third one a message $message_n$. We check the equality between actor's address ($X_1^2$) and receiver ($X_1^3$),
behavior label ($X_1^1$) and message label ($X_2^3$). With the $behavior\_set$ function we check the link between the behavior
and the actor. The value passing is defined in the same way as in the first rule.
%\begin{figure}[h!]
\begin{center}
\(
dynamic\_trans_n = (3, components , compatibility , v\_passing )
\)
\end{center}
where\\
\begin{tabular}{l l}
%\begin{enumerate}
%\item 

1. $components  = \left\{ 
  \begin{array}{l c l} 
    1 &\mapsto& behavior_n,\\ 
    2 &\mapsto& dynamic\_actor, \\ 
    3 & \mapsto& message_n
  \end{array}
  \right.$
&
%\item 

2. $compatibility = \left\{ \begin{array}{l}
  X_1^2 = X_1^3;\\
  behavior\_set (I^1) = X_2^2;\\
  X_1^1 = X_2^3 ; \\
\end{array} \right. $

\\ \\
%\item 

3. $v\_passing = \left\{ \begin{array}{l}
Y_1^1 \leftarrow X_1^2;\\
Y_2^1 \leftarrow X_2^2;\\
Y_{i+2}^1 \leftarrow X_{i+2}^3, \forall i \in \llbracket 1;n \rrbracket;
\end{array}
\right.$

%\end{enumerate}
\end{tabular}
%\caption{Communication with a dynamic actor}
%\label{rule_2}
%\end{figure}

\subsection{Operational semantics}

We now briefly describe how to use the preceding definitions to express in the non standard syntax both an initial term and 
the computation of a transition according to a formal rule.

Initial configurations are obtained by launching a continuation in $init_s$ with an empty marker and an empty environment. 
That means inserting in an empty configuration, one thread for each pair $(p,E_s)$ in $\beta(init_s)$ where each value in $E_s$ is 
associated with an empty marker. We focus now on the interaction computation according to one of the two rules. 
First of all, we have to find some correct interaction. It means that we have to find some threads in the current configuration 
that can be associated to the right partial interaction according to the matching formal rule. 
Then we check that their interface satisfies the synchronization constraints. Thus we can compute
the interaction:
\begin{itemize}
\item we remove interacting threads according to the type of their exhibited partial interaction;
\item we choose a syntactic continuation for each thread;
\item we compute dynamic data for each of these continuations:
  \begin{itemize}
  \item we compute the marker;
  \item we take into account name passing;
  \item we create fresh variables and associate them with the correct values;
  \item we restrict the environment according to the interface associated with the program point.
  \end{itemize}
\end{itemize}

\subsection{Correspondence}
\begin{thm}[correspondence]
CAP standard semantics and its non standard semantics are in strong bisimulation
\end{thm}

\begin{proof}
The proof can be found at the first author's web page, www.enseeiht.fr/~garoche.
\end{proof}

\subsection {Example}

To illustrate the use of the non standard semantics, we will compute the first transition of the example given in 
section~\ref{cap}.

The initial configuration\footnote{We can notice the absence of threads at program points $3$, $5$ and $9$ which correspond to sub-terms. There are not present in the initial configuration.} is:
\begin{center}
\(
\begin{array}{c}
(1,\epsilon,\left[\begin{array}{l c l} a &\mapsto& \alpha,\epsilon\end{array}\right]) \quad
(2,\epsilon,\left[\begin{array}{l c l} a &\mapsto& \alpha,\epsilon \end{array}\right]) \quad
(4,\epsilon,\left[\begin{array}{l c l} \end{array}\right]) \quad 
(6,\epsilon,\left[\begin{array}{l c l} a &\mapsto& \alpha,\epsilon\\ b &\mapsto& \beta,\epsilon \end{array}\right]) \\
(7,\epsilon,\left[\begin{array}{l c l} b &\mapsto& \beta,\epsilon\end{array}\right] ) \quad
(8,\epsilon,\left[\begin{array}{l c l}\end{array}\right]) \quad
(10,\epsilon,\left[\begin{array}{l c l} b &\mapsto& \beta,\epsilon\end{array}\right])
\end{array}
\)
\end{center}

At this point, the only possible transition is labeled by $1,6$ and corresponds to the $static\_trans_n$ rule.
Program point $1$ is able to exhibit the two following partial interactions:
%\begin{center}
\( \left\{ \begin{array}{l}
\Big\{ ( static\_actor_n , [a,m], [e,s], \beta(a \act^3 s,\emptyset) ) \Big\} , \\
\Big\{ ( static\_actor_n , [a,send], [e,s,x], \beta(x \mess^5 beh(s),\emptyset) ) \Big\} 
\end{array}
\right\}
\) %\end{center}
when the program point $6$ exhibits the only partial interaction:
\begin{center}
\( \Big\{ ( message_n , [a,send,b] , \emptyset , \emptyset ) \Big\} \)
\end{center}

We choose the first partial interaction for $1$. We first check synchronization constraints. 
We need that $X_1^1 = X_1^2$ and $X_2^1 = X_2^2$. So $(\alpha,\epsilon)= (\alpha, \epsilon)$ and 
both message share the same label $send$. We can now compute value passing, thread launching and removing.
We have to remove interacting threads and to add threads in $\beta(x \mess^5 beh(s),\emptyset)$ with their environment updated
by value passing. Value passing gives the value of $e$, $s$ and $x$, we have respectively, $(\alpha,\epsilon)$, $(1,\epsilon)$ and
$(\beta,\epsilon)$. Thus the launched thread is $(5,\epsilon,\left[\begin{array}{l c l} x &\mapsto& \beta,\epsilon\\s & \mapsto& 1,\epsilon\end{array}\right])$.

We obtain the new configuration:
\begin{center}
\(
\begin{array}{l l l}
(2,\epsilon,\left[\begin{array}{l c l} a &\mapsto& \alpha,\epsilon \end{array}\right])&
(4,\epsilon,\left[\begin{array}{l c l} \end{array}\right]) &
(5,\epsilon,\left[\begin{array}{l c l} x &\mapsto& \beta,\epsilon\\s & \mapsto & 1, \epsilon \end{array}\right]) \\ \\
(7,\epsilon,\left[\begin{array}{l c l} b &\mapsto& \beta,\epsilon \end{array}\right] ) &
(8,\epsilon,\left[\begin{array}{l c l} \end{array}\right])&
(10,\epsilon,\left[\begin{array}{l c l} b &\mapsto& \beta,\epsilon\end{array}\right])\\
\end{array}
\)
\end{center}

We recall that when computing a transition using the $dynamic\_trans_n$ rule, new launched threads are associated to a new marker.

\section{Abstract semantics}

In order to ensure properties on all the possible execution of the non standard semantics, we rely on the abstract interpretation 
approach which combines in a single one all the possible executions.

\subsection{Abstract Interpretation}

Abstract interpretation~\cite{CousotCousot77-1} is a theory of discrete approximation of semantics. 
A fundamental aspect of this theory is
that every semantics can be expressed as fixed points of monotonic operators on complete partial orders. 
A concrete semantics is defined by a tuple $(S,\subseteq,\bot,\cup,\top,\cap)$.
Following~\cite{CousotCousot92-2}, an abstract semantics is defined by a pre-ordered set $(S^\#,\sqsubseteq)$, 
an abstract iteration basis $\bot^\#$,
a concretization function $\gamma: S^\# \rightarrow S$ and an
abstract semantics function $\mathbb{F}^\#$.

\subsubsection{Abstract interpretation of mobile systems.}
We approximate here the mobile systems' semantics as described in~\cite{feret_eng:phd,venet:phd}. 
The collecting semantics of a configuration $\mathscr{C}_0$ is defined as the least fixed point of the complete join morphism 
$\mathbb{F}$:
\begin{center}
\(
\mathbb{F}(X) = ( \{ \epsilon \} \times \mathscr{C}_0 ) \cup
\left\{ \begin{array}{l | l}
(u.\lambda,C') &
\exists C \in \mathscr{S}, (u,C) \in X \textrm{ and } C \xrightarrow{\lambda} C'
\end{array} \right\} 
\)
\end{center}
 An abstraction $\left( \mathscr{C}^\# , \sqsubseteq^\#,
 \sqcup^\#, \bot^\#, \right.$ $\left. \gamma^\#, C_0^\#, \rightsquigarrow , \nabla \right)$
in this framework must define as usual a pre-order, a join operator, a bottom element, a widening operator (when abstract 
domains are infinite) as well as:
\begin{itemize}
\item the initial abstract configuration 
  $C_0^\# \in \mathscr{C}^\#$ with $\{ \epsilon \} \times \mathscr{C}_0 \subseteq \gamma(C_0^\#)$
\item the abstract transition relation $\rightsquigarrow ~ \in \wp (\mathscr{C}^\# \times \Sigma \times \mathscr{C}^\#)$  
  such that:

  $\forall C^\# \in \mathscr{C}^\#, \forall (u,C) \in \gamma(C^\#),
  \forall \lambda \in \Sigma, \forall C' \in \mathscr{C}$,
  \begin{center}
  \(
  C \xrightarrow{\lambda} C' \implies \exists C'^\# \in \mathscr{C}^\#, (C^\# \stackrel{\lambda}{\rightsquigarrow} C'^\#) 
  \textrm{ and } (u.\lambda, C') \in \gamma(C'^\#)
  \)
\end{center} 
Such an abstract transition computes all the concrete transitions labeled $\lambda$ from all possible $C$ represented by $C^\#$.
\end{itemize}

The abstract counterpart of the $\mathbb{F}$ function is the abstract function $\mathbb{F}^\#$ defined as:
\begin{center}
\(
\mathbb{F}^\# (C^\#) = {\bigsqcup}^\# \left( \left\{ C'^\# ~|~ \exists \lambda \in \Sigma, C^\# \rightsquigarrow^\lambda C'^\# \right\} 
\sqcup \{ C_0^\#; C^\# \}
\right)
\)
\end{center}

\subsection{Abstract Domains}
An element of an abstract domain expresses the set of invariant properties of a set of terms. 
We project the initial term into an abstract element to
describe its properties. Then we use an abstract counterpart of the transition rules to obtain the set of valid properties 
when applying the transition rule to all elements of the initial set.
Then we compute the union of both abstract elements, to only keep the set of properties which
are valid before and after the transition. 
We repeat these steps until a fixed point is reached. The use of the union and the widening functions guarantees the monotony of the transition and thus
the existence of the fixed point. Finally, we obtain an abstract element describing
the set of valid properties in all possible evolutions of the initial term. It is a post fixed point of the collecting semantics' 
least fixed point. Our abstractions are sound counterparts of the non standard semantics. 

In order to avoid a too coarse approximation of the collecting semantics, we need, at least, to use a good abstraction of the 
control flow. We associate to each program point an abstract element describing its set of values and markers. 
But, most of our properties  
can be expressed in terms of occurrence counting. We also need to approximate configurations globally.
Therefore, we use, as an abstract domain, the cartesian product of an abstract domain to approximate non uniform control flow 
information in conjunction with a domain to approximate the occurrence of threads in configurations.

\subsubsection{Generic abstractions.}
In this section, we will briefly describe the two abstract domains defined, by \textsc{Feret}, respectively in~\cite{feret:esop02} 
and~\cite{feret:occurrence-counting} that are used
to approximate the non standard semantics of CAP. Their operational semantics is then given in Figs.~\ref{sem_op_flow}
and~\ref{sem_op_occ}.

\paragraph{Control Flow Abstract Domain.}
This abstract domain approximates variable values of thread environments as well as their marker for a given configuration. 
It is parametrized by 
an abstract domain called an Atom Domain. We associate to each program point an atom which describes the values of both variables
and markers of the threads that can be associated with this program point. When computing an interaction, we merge the interacting
atoms associated to the interacting threads (primitive $reagents^\#$) and add synchronization constraints (primitive $sync^\#$).
If they are satisfiable, the interaction is possible. We then compute the value passing and the marker computation 
(function $marker\_value$). Finally, we launch new threads (primitive $launch^\#$) and update the atom of each program point 
by computing its union with the appropriate resulting atom. 

In this domain, we only focus on values, so we completely abstract away occurrences of threads and thus
deletion of interacting threads.

The Atom Domain we use is a reduced product of four domains. The first two represent equality and disequality among
values and marker using graphs, the third one approximates the shape of markers and values with an automaton
and the fourth one approximates the relationship between occurrences of letters in Parikh's vectors~\cite{parikh}
associated to each value and marker.

\paragraph{Occurrence Counting Abstract Domain.}
In this domain, we count both threads associated to a particular program point and transition label, 
the set of which is denoted by $\mathscr{V}_c$. 
We first approximate the non standard semantics by the domain $\mathbb{N}^{\mathscr{V}_c}$ 
associating to each program point its threads occurrence in the configuration and to each 
transition label, its occurrence in the word that leads to the configuration. 
At the level of the collecting semantics, we obtain an element in $\wp(\mathbb{N}^{\mathscr{V}_c})$. We then abstract 
such a domain by a domain $\mathscr{N}_{\mathscr{V}_c}$ which is 
a reduced product between the domain of intervals indexed by $\mathscr{V}_c$ and the domain of affine equalities~\cite{karr} 
constructed over $\mathscr{V}_c$. When computing a transition, we check that the occurrences of interacting threads are sufficient to 
allow it (primitive $SYNC_{\mathscr{N}_{\mathscr{V}_c}}$). If we do not obtain the bottom element of our abstract domain, \ie 
the synchronization constraint is satisfiable, we add 
(primitive $+^\#$) the new transition label, the launched threads (primitives $\beta^\#$ and $\Sigma^\#$) and remove 
(primitive $-^\#$) consumed threads.

\begin{figure}[t]
Let $C^\#$ be an abstract configuration,
let $(p_k)_{1 \leq k \leq n} \in \mathscr{L}_p$ be a tuple of program points label and
$(pi_k)_{1 \leq k \leq n} = (s_k, (parameter_k),(bd_k),constraints_k,continuation_k)$ be a tuple of partial interactions.

\subfigure[Abstract semantics for control flow approximation.]{
\label{sem_op_flow}
\begin{tabular}{l}
We define $mol$ by $reagents^\# ((p_k),(parameter_{k,l}), (constraints_k) , C^\# )$.\\
When\\
\begin{tabular}{l}
\( \forall k \in \llbracket 1;n \rrbracket, pi_k \in interaction(p_k) \) ;\\
\( mol \neq  \bot_{(I(p_k))_k}\)
\end{tabular}\\
Then\\
\multicolumn{1}{c}{$C {\xrightarrow{(p_k)_k}}_\# \bigsqcup \{ C ; mol; new\_threads \}$}\\
Where\\
\begin{tabular}{l}
1.~\( mol' = marker\_value((p_k)_k, mol, (bd_{k,l})_{k,l}, (parameter_{k,l})_{k,l}, \) \(\mathnormal{v\_passing} )\)\\
2.~$new\_threads = launch^\# ((p_k, continuations_k)_k, mol')$.
\end{tabular}
\end{tabular}
}
\subfigure[Abstract semantics for occurrence counting.]{
\label{sem_op_occ}
\begin{tabular}{l}
We define the tuple $t \in \mathbb{N}^{\mathscr{V}_c}$ so that
$t_v$ be the occurrence of $v$ in $(p_k)_{1 \leq k \leq n}$.\\
When\\
\begin{tabular}{l}
\( \forall k \in \llbracket 1;n \rrbracket, pi_k \in interaction(p_k) \) ;
\( SYNC_{\mathscr{N}_{\mathscr{V}_c}} (t,C^\#) \neq \bot_{\mathscr{N}_{\mathscr{V}_c}} \)
\end{tabular}\\
Then\\
\multicolumn{1}{c}{$C {\xrightarrow{(p_k)_k}}_\# SYNC_{\mathscr{N}_{\mathscr{V}_c}} (t,C^\#) +^\# Transition +^\# Launched -^\# Consumed$}
\\
Where\\
\begin{tabular}{l}
1.~$Transition = 1_{\mathscr{N}_{\mathscr{V}_c}} (p_1)$;\\
2.~$Launched = \Sigma^\# \left( (\beta^\# (continuation^k))_k\right)$;\\
3.~$Consumed = {\Sigma^\# (1_{\mathscr{N}_{\mathscr{V}_c}} (p_k))_{k \in \{ k' | 1 \leq k' \leq n, type(s_{k'}) \neq replication\} } }$
\end{tabular}
\end{tabular}
}
\caption{Abstract operational semantics}
\label{abs_sem}
\end{figure}

\section{Properties}
The abstract semantics computes an approximation of all the execution 
in the non standard one. Its result can then be used in 
order to check many different properties. In this section, we describe 
interesting properties and how to observe them in the fixed point of 
the analysis.

\subsection{Linearity}

Linearity is a property that expresses the fact that all actors in each possible configuration are bound to different addresses.
It can be expressed as in $\pi$-calculus when each process listens to at most one channel. It is a useful property
to map addresses to resources.

Our analysis is able to prove that a term, without recursive name definitions, \ie without a $\nu$ operator inside 
a behavior continuation, will be linear in all the 
possible configurations it will take. We can observe
such a property with both the control flow domain and the occurrence counting domain.
We first determine with the control flow the upper set of program points representing actors
that can be associated with each address. Then we check in the occurrence counting domain that each of those
program points is mapped to at most one thread in each configuration (within the interval domain) and, moreover, 
that program points
that can be associated with the same address are in mutual exclusion (with the global numerical domain).
The mutual exclusion property is observed by exhibiting a constraint from the global numerical domain. Such a constraint must
be a linear combination $\Sigma x_i + \Sigma k_j * y_j = 1$ with $\{ x_i \}$ the set of 
program points in mutual exclusion and $\{ k_j \}$ a set of positive or null coefficients. Whether such a constraint can be generated
by the set of constraints describing the affine space of the global numerical domain then the $\{ x_i \}$ program points are in
mutual exclusion but they do not have to be present in every configuration of the system.

In the following example, we can automatically determine that the following term satisfies the linearity property.
\begin{center}
\(
\begin{array}{l l}
\nu a^\alpha, b^\beta,&  a \act^{1:\llbracket 0;1 \rrbracket} [m()^{2:\llbracket 0;1 \rrbracket}() = \zeta(e,s)(e \act^{3:\llbracket 0;1 \rrbracket} s),\\
& ~~~~~~~~~~~~~ send^{4:\llbracket 0;1 \rrbracket}(x) = \zeta(e,s)(x \mess^{5:\llbracket 0;1 \rrbracket} beh(s))] \\
~||~ & b \act^{6:\llbracket 0;1 \rrbracket} [beh^{7:\llbracket 0;1 \rrbracket}(x) = \zeta(e,s)(e \act^{8:\llbracket 0;1 \rrbracket} x)]\\
~||~ & a \mess^{9:\llbracket 0;1 \rrbracket} send(b) ~||~ b \mess^{10:\llbracket 0;1 \rrbracket} m()
\end{array}
\) 
\end{center}

All the actors are associated with the interval $\llbracket 0;1 \rrbracket$. The only actor that can be associated to address $a$
is $1$ and others ($3,6$ and $8$) can be associated with address $b$. Then the constraint $p_3 + p_6 + p_8 =1$ can be observed
in the global numerical part of the post fixed point of the analysis. We can notice that we have a stronger 
property: there is exactly 
one actor on the address $b$ in every configuration of this term.

\subsection{Bounded resources}
As CAP is an asynchronous calculus, when a message is sent we cannot ensure that it will be handled.
With this property, we want to determine if the system grows infinitely; if the system creates more messages than it can handle.
Our analysis is able to infer such a property. We first check which message can have an unbounded number of occurrences.
Then we check in the global numerical invariants of the system a constraint between the number of occurrences of this message and
the number of occurrences of a transition labeled with the same message label. 
When such a constraint can be found, we can say that this message will be
in the system an unbounded number of times, but it will be handled the same number of times. The system size is constant, it does not diverge.

In the following example, our analysis is able to find that at most one message is present in the system: program points $3$, $7$ and $9$ associated with interval $\llbracket 0;1 \rrbracket$. The system described by this term is bounded. 
Furthermore, we have the constraint $p_3 + p_7 + p_9 = 1$. 
\begin{center}
\(
\begin{array}{r l}
\nu a^\alpha, \nu b^\beta, & a \act^{1:\llbracket 0; 1 \rrbracket} [ ping^{2:\llbracket 1; 1 \rrbracket}() = \zeta(e, s) ( b \mess^{3:\llbracket 0; 1 \rrbracket} pong() ~||~ e \act^{4:\llbracket 0; 1 \rrbracket} s)]  \\
|| & b \act^{5:\llbracket 0; 1 \rrbracket} [ pong^{6:\llbracket 1; 1 \rrbracket}() = \zeta(e, s) ( a \mess^{7:\llbracket 0; 1 \rrbracket} ping() ~||~ e \act^{8:\llbracket 0; 1 \rrbracket} s)]\\
||& a \mess^{9:\llbracket 0; 1 \rrbracket} ping()
\end{array}
\)
\end{center}

In addition, we can also detect whether a system does not generate an unbounded number of actor present at the same time in 
a given configuration. 
\begin{center}
\(
\nu a^\alpha a \act^{1:\llbracket 0; 1 \rrbracket} [ m^{2:\llbracket 1; 1 \rrbracket}() = \zeta(e, s) ( \nu b^\beta b \act^{3:\llbracket 0; 1 \rrbracket} s ~||~ b \mess^{4:\llbracket 0; 1 \rrbracket} m())]  ~||~ {a \mess^{5:\llbracket 0; 1 \rrbracket} m()}
\)
\end{center}

In the preceding example, we automatically detect that the number of threads associated to program point $3$ lies in
$\llbracket 0; 1 \rrbracket$.

\subsection{Unreachable behaviors}

We are interested in determining
the subset of behaviors that are really used for each set of behaviors.
Due to its high-order capability, CAP allows to send the set of behaviors
syntactically associated to an actor to other actors. Therefore the use of the
behavior's set depends highly on the messages exchanged.

In the following example, all the behavior branches of the behavior syntactically defined at program point $1$ are used.
We check such a property by checking that each label of transition is present at least once or its continuation has been 
launched. \Ie $\forall t \in \mathscr{V}_c, Inter(t) \neq \llbracket 0; 0\rrbracket$ where $Inter$ is the function that maps
each element of $\mathscr{V}_c$ to its image in interval part of the analysis post fixed point.

\begin{center}
\(
\begin{array}{r l}
\nu a^\alpha, b^\beta, c^\gamma, &
a \act^1 [ m_0^2() = \zeta(e,s)( b \mess^3 n_1(s) ~||~ b \mess^4 m_1(c) ), \\
& ~~~~~~~~m_1^5(dest) = \zeta(e,s)( dest \mess^6 m_2() ), \\
& ~~~~~~~~m_2^7() = \zeta(e,s)(\emptyset) ] \\
 || &b \act^8 [n_1^9(self) = \zeta(e,s)( e \act^{10} self ~||~ c \mess^{11} n_2(self)) ]\\
 || &c \act^{12} [n_2^{13}(self) = \zeta(e,s)( e \act^{14} self)] \\
 || &a \mess^{15} m_0()
\end{array}
\)
\end{center}

We can use such an analysis to clean the term with garbage collecting like mechanisms.

\section{Conclusion}

We have adapted the framework of \textsc{Feret}~\cite{feret_eng:phd} to deal 
with a higher order process calculus modeling actor languages. 
With such
a framework, we are able to analyze CAP terms without any restriction about the kind of values sent within messages: we can
now handle behavior passing, which was not able with our previous type based analysis. 
In contrary to our aforementioned analyses about actor's calculus, we are able to easily count occurrences of 
both actors and messages. Therefore, most of the properties we obtain are related to occurrence counting. We can detect whether
the number of actors and messages is finite, whether there is dead code and whether the message queues are bounded. 
We also have the linearity property under certain restrictions.

To go further, we need another abstraction which will split thread's information into computation units representing the 
recursive instances of the same thread. Such an abstract domain will allow us to deal with linearity in the general case as well
as handling more properties. In fact the most interesting property with an asynchronous process calculus with non uniform behavior, 
is the detection of orphan messages, \ie stuck-freeness. 
An orphan is a message which may not be handled by its target in some execution path. We distinguish two kinds of orphan: safety
ones and liveness ones. Safety orphans occur when all future behaviors of the target on a given execution path cannot 
handle such a message.
On the contrary, liveness orphans occur when one of the target behaviors in each execution paths knows how to handle
such a message but the target is deadlocked and will never assume the corresponding behavior. We advocate that with this new 
abstract domain we will be able to detect both kinds of orphans. We also want to define a generic abstract domain dedicated to
the data-flow like analyses provided by type systems. Such an abstract domain can be useful to automatically build domains 
to observe properties for which we already have a type system. 

\begin{ack}
We deeply thank Jérome Feret for fruitful discussions and careful proof 
reading of the first author's Master's thesis~\cite{garoche:master}.
\end{ack}

\bibliography{biblio/biblio_lima,biblio/biblio_ens,biblio/biblio}
\bibliographystyle{plain}

\end{document}